\documentclass[reprint,amsmath,amssymb,jap,aip,onecolumn]{revtex4-1}
\usepackage{graphicx}
\usepackage{dcolumn}
\usepackage{bm}
\usepackage{epsfig}
\usepackage{mathptmx}
\usepackage{times}
\usepackage{amsmath}
\usepackage{amssymb}
\usepackage{units}
\usepackage{upgreek}
\usepackage{float}
\usepackage{color}
\usepackage{soul}

\newcommand{\Iprog}{$I_\text{prog}$}

\newcommand{\GST}{Ge$_2$Sb$_2$Te$_5$}

\begin{document}
\noindent
\large 
This manuscript was published in: Nature Communications, volume 9, page 2514 (2018) \\
Link to published article: https://www.nature.com/articles/s41467-018-04933-y \\
DOI: 10.1038/s41467-018-04933-y \\

\normalsize 

\title{Neuromorphic computing with multi-memristive synapses}

\author{Irem Boybat}\email{ibo@zurich.ibm.com} \affiliation{IBM Research -- Zurich, S\"{a}umerstrasse 4, 8803 R\"{u}schlikon, Switzerland.} \affiliation{Microelectronic Systems Laboratory, EPFL, Bldg ELD, Station 11, CH-1015 Lausanne, Switzerland}
\author{Manuel Le Gallo} \affiliation{IBM Research -- Zurich, S\"{a}umerstrasse 4, 8803 R\"{u}schlikon, Switzerland.}
\author{S. R. Nandakumar} \affiliation{IBM Research -- Zurich, S\"{a}umerstrasse 4, 8803 R\"{u}schlikon, Switzerland.} \affiliation{Department of Electrical and Computer Engineering, New Jersey Institute of Technology, Newark, NJ 07102, USA}
\author{Timoleon Moraitis} \affiliation{IBM Research -- Zurich, S\"{a}umerstrasse 4, 8803 R\"{u}schlikon, Switzerland.}
\author{Thomas Parnell} \affiliation{IBM Research -- Zurich, S\"{a}umerstrasse 4, 8803 R\"{u}schlikon, Switzerland.}
\author{Tomas Tuma} \affiliation{IBM Research -- Zurich, S\"{a}umerstrasse 4, 8803 R\"{u}schlikon, Switzerland.}
\author{Bipin Rajendran} \affiliation{Department of Electrical and Computer Engineering, New Jersey Institute of Technology, Newark, NJ 07102, USA}
\author{Yusuf Leblebici} \affiliation{Microelectronic Systems Laboratory, EPFL, Bldg ELD, Station 11, CH-1015 Lausanne, Switzerland}
\author{Abu Sebastian}\email{ase@zurich.ibm.com} \affiliation{IBM Research -- Zurich, S\"{a}umerstrasse 4, 8803 R\"{u}schlikon, Switzerland.}
\author{Evangelos Eleftheriou} \affiliation{IBM Research -- Zurich, S\"{a}umerstrasse 4, 8803 R\"{u}schlikon, Switzerland.}

\begin{abstract}
Neuromorphic computing has emerged as a promising avenue towards building the next generation of intelligent computing systems. It has been
proposed that memristive devices, which exhibit history-dependent conductivity modulation, could efficiently represent the synaptic weights in
artificial neural networks. However, precise modulation of the device conductance over a wide dynamic range, necessary to maintain high network
accuracy, is proving to be challenging. To address this, we present a multi-memristive synaptic architecture with an efficient global
counter-based arbitration scheme. We focus on phase change memory devices, develop a comprehensive model and demonstrate via
simulations the effectiveness of the concept for both spiking and non-spiking neural networks. Moreover, we present experimental results involving
over a million phase change memory devices for unsupervised learning of temporal correlations using a spiking neural network. The work presents a significant step
towards the realization of large-scale and energy-efficient neuromorphic computing systems.
\end{abstract}
\maketitle

The human brain with less than 20 Watts of power consumption offers a processing capability that exceeds the petaflops mark and thus outperforms
state-of-the-art supercomputers by several orders of magnitude in terms of energy efficiency and volume. Building ultra-low-power cognitive computing
systems inspired by the operating principles of the brain is a promising avenue towards achieving such efficiency. Recently, deep learning has
revolutionized the field of machine learning by providing human-like performance in areas such as computer vision, speech recognition, and complex
strategic games \cite{Y2015lecunNature}. However, current hardware implementations of deep neural networks are still far from competing with
biological neural systems in terms of real-time information-processing capabilities with comparable energy consumption.

One of the reasons for this inefficiency is that most neural networks are implemented on computing systems based on the conventional von
Neumann architecture with separate memory and processing units. There are a few attempts to build custom neuromorphic hardware that is optimized to
implement neural algorithms \cite{Y2010shemmelISCAS, Y2013painkrasSolid, Y2014merollaScience, Y2015qiaoFNINS}. However, as these custom systems are
typically based on conventional silicon CMOS circuitry, the area efficiency of such hardware implementations will remain relatively low, especially
if \textit{in-situ} learning and non-volatile synaptic behaviour have to be incorporated. Recently, a new class of nanoscale devices has
shown promise for realizing the synaptic dynamics in a compact and power-efficient manner. These memristive devices store information in their
resistance/conductance states and exhibit conductivity modulation based on the programming history \cite{Y2000beckAPL, Y2008strukovNature,
Y2011chuaAPA, Y2015wongNatureNano}. The central idea in building cognitive hardware based on memristive devices is to store the synaptic weights as
their conductance states and to perform the associated computational tasks in place.

The two essential synaptic attributes that need to be emulated by memristive devices are the synaptic efficacy and plasticity. Synaptic efficacy
refers to the generation of a synaptic output based on the incoming neuronal activation. In conventional non-spiking artificial neural
networks (ANN), the synaptic output is obtained by multiplying the real-valued neuronal activation with the synaptic weight. In a spiking
neural network (SNN), the synaptic output is generated when the presynaptic neuron fires and typically is a signal that is proportional to the
synaptic conductance. Using memristive devices, synaptic efficacy can be realized using Ohm's law by measuring the current that flows through the
device when an appropriate read voltage signal is applied. Synaptic plasticity, in contrast, is the ability of the synapse to change its weight,
typically during the execution of a learning algorithm. An increase in the synaptic weight is referred to as potentiation and a decrease as depression. In an ANN, the weights are usually changed based on the backpropagation
algorithm \cite{1986rumelhartParallel}, whereas in an SNN, local learning rules such as spike-timing-dependent plasticity (STDP)
\cite{Y1997markramScience} or supervised learning algorithms such as NormAD \cite{Y2015anwaniIJCNN} could be used. The implementation of synaptic
plasticity in memristive devices is achieved by applying appropriate electrical pulses that change the conductance of these devices through
various physical mechanisms \cite{Y2015saighiFNINS, Y2016burrAdvX, Y2016rajendranJETCAS}, such as ionic drift \cite{Y2011ohnoNatMat, Y2011yuITED,
Y2016ambrogioITED, Y2016coviFrontiers, Y2017burgtNatMat}, ionic diffusion \cite{Y2017wangNatMat}, ferroelectric effects \cite{Y2017boynNatComm},
spintronic effects \cite{Y2012wuJETCAS, Y2015VincentBiomedical} and phase transitions \cite{Y2012kuzumNano, Y2016ambrogioFNINS}.

Demonstrations that combine memristive synapses with digital or analog CMOS neuronal circuitry are indicative of the potential to realize highly
efficient neuromorphic systems \cite{Y2013alibartNatComm, Y2013indiveriNanotech, Y2015preziosoNature, Y2015kimIEDM, Y2015mostafaFNINS, Y2016tumaEDL,
Y2016wozniakISCAS}. However, to incorporate such devices into large-scale neuromorphic systems without compromising the network performance,
significant improvements in the characteristics of the memristive devices are needed\cite{Y2015burrITED}. Some of the device characteristics that limit the
system performance include the limited conductance range, asymmetric conductance response (differences in the manner in which the conductance changes
between potentiation and depression), nonlinear conductance response (nonlinear conductance evolution with respect to the number of pulses),
stochasticity associated with conductance changes, and variability between devices.

Clearly, advances in materials science and device technology could play a key role in addressing some of these challenges
\cite{Y2015koelmansNatComm,fuller2017li}, but equally important are innovations in the synaptic architectures. One example is the differential synaptic
architecture \cite{Y2011suriIEDM}, in which two memristive devices are used in a differential configuration such that one device is used for
potentiation and the other for depression. This was proposed for synapses implemented using phase change memory (PCM) devices which exhibit strong
asymmetry in their conductance response. However, the device mismatch within the differential pair of devices as well as the need to refresh the
device conductance frequently to avoid conductance saturation could potentially limit the applicability of this approach \cite{Y2015burrITED}. In
another approach proposed recently \cite{Y2014billFN}, several binary memristive devices are programmed and read in parallel to implement a synaptic
element, exploiting the probabilistic switching exhibited by certain types of memristive devices. However, it may be challenging to achieve fine-tuned probabilistic switching
reliably across a large number of devices. Alternatively, pseudo-random number generators
could be used to implement this probabilistic update scheme with deterministic memristive devices \cite{Y2015garbinIEDM}, albeit with the associated
costs of increased circuit complexity.

In this article, we propose a multi-memristive synaptic architecture that addresses the main drawbacks of the above-mentioned schemes, and experimentally demonstrate an implementation using nanoscale PCM devices. First, we present the concept of multi-memristive synapses with a
counter-based arbitration scheme. Next, we illustrate the challenges posed by memristive devices for neuromorphic computing by studying the operating
characteristics of PCM fabricated in the \unit[90]{nm} technology node and show how multi-memristive synapses can address some of these challenges. Using
comprehensive and accurate PCM models, we demonstrate the potential of the multi-memristive synaptic concept in training ANNs and SNNs for the
exemplary benchmark task of handwritten digit classification. Finally, we present a large-scale experimental implementation of training an SNN with
multi-memristive synapses using more than one million PCM devices to detect temporal correlations in
event-based data streams.

\section{Results}
\subsection{The multi-memristive synapse}\label{sec:multi}
\begin{figure*}[t!]
\centering
\begin{tabular}{c}
\includegraphics[scale = 1]{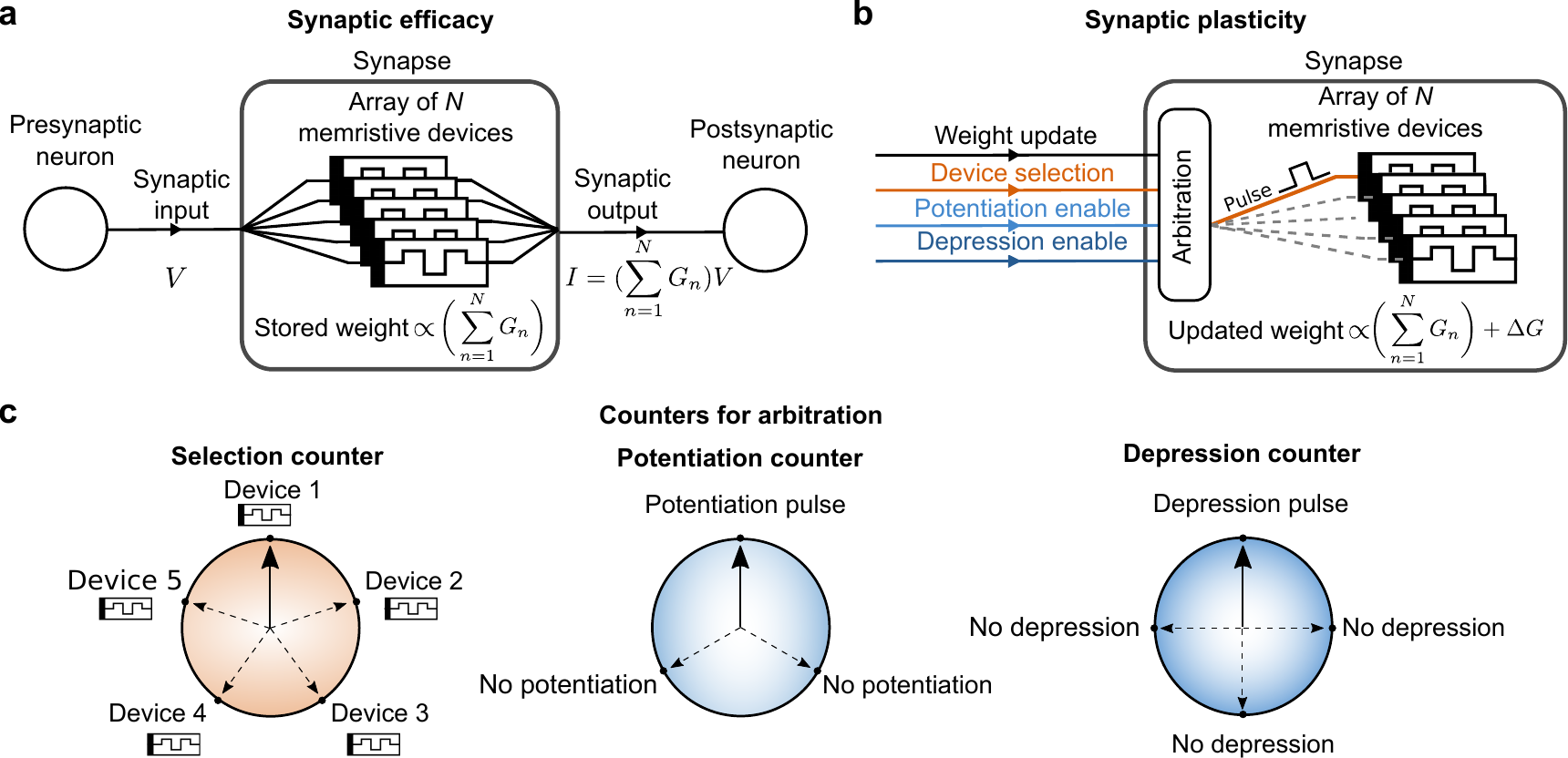}
\end{tabular}
\caption{\textbf{The multi-memristive synapse concept.} (a) The net synaptic weight of a multi-memristive synapse is represented by the combined
conductance ($\sum G_n$) of multiple memristive devices. To realize synaptic efficacy, a read voltage signal, $V$, is applied to all devices. The
resulting current flowing through each device is summed up to generate the synaptic output. (b) To capture synaptic plasticity, only one of the
devices is selected at any instance of synaptic update. The synaptic update is induced by altering the conductance of the selected device as dictated
by a learning algorithm. This is achieved by applying a suitable programming pulse to the selected device. (c) A counter-based arbitration scheme is
used to select the devices that get programmed to achieve synaptic plasticity.  A global selection counter whose maximum value is equal to the number
of devices representing a synapse is used. At any instance of synaptic update, the device pointed to by the selection counter is programmed.
Subsequently, the selection counter is incremented by a fixed amount. In addition to the selection counter, independent potentiation and depression
counters can serve to control the frequency of the potentiation or depression events.} \label{fig:multimem}
\end{figure*}

The concept of the multi-memristive synapse is illustrated schematically in Fig. \ref{fig:multimem}. In such a synapse, the synaptic weight is
represented by the combined conductance of $N$ devices. By using multiple devices to represent a synaptic weight, the overall dynamic range and
resolution of the synapse are increased. For the realization of synaptic efficacy, an input voltage corresponding to the neuronal activation is
applied to all constituent devices. The sum of the individual device currents forms the net synaptic output. For the implementation of synaptic
plasticity, only one out of $N$ devices is selected and programmed at a time. This selection is done with a counter-based arbitration scheme where
one of the devices is chosen according to the value of a counter (see Supplementary Note 1). This selection counter takes
values between 1 and $N$, and each value corresponds to one device of the synapse. After the weight update, the counter is incremented by a fixed
increment rate. Having an increment rate co-prime with the clock length $N$ guarantees that all devices in each synapse will eventually get selected
and will receive a comparable number of updates provided there is a sufficiently large number of updates. Moreover, if a single selection clock is
used for all synapses of a neural network,  $N$ can be chosen to be co-prime with the total number of synapses in the network to avoid updating the
same device in a synapse repeatedly.

In addition to the global selection counter, additional independent counters, such as a potentiation counter or a depression
counter, could be incorporated to control the frequency of potentiation/depression events (see Fig. \ref{fig:multimem}). The value of the
potentiation (depression) counter acts as an enable signal to the potentiation (depression) event; a potentiation (depression) event is enabled if
the potentiation (depression) counter value is one, and is disabled otherwise (see Supplementary Note 2). The frequency of the potentiation
(depression) events is controlled by the maximum value or length of the potentiation (depression) counter. The counters are incremented after the
weight update. By controlling how often devices are programmed for a conductance increase or decrease, asymmetries in the device conductance response
can be reduced.

The constituent devices of the multi-memristive synapse can be arranged in either a differential or a non-differential architecture. In the
latter each synapse consists of $N$ devices, and one device is selected and potentiated/depressed to achieve synaptic
plasticity. In the differential architecture, two sets of devices are present, and the synaptic conductance is calculated as $G_\text{syn} = G_+ -
G_-$, where $G_+$ is the total conductance of the set representing the potentiation of the synapse and $G_-$ is the total conductance of the set
representing the depression of the synapse. Each set consists of $N/2$ devices. When the synapse has to be potentiated, one device from the group
representing $G_+$ is selected and potentiated, and when the synapse has to be depressed, one device from the group representing $G_-$ is selected
and potentiated.

An important feature of the proposed concept is its crossbar compatibility. In the non-differential architecture, by placing the devices
that constitute a single synapse along the bit lines of a crossbar, it is possible to sum up the currents using Kirchhoff's law and obtain the total
synaptic current without the need for any additional circuitry (see Supplementary Note 3).  The differential architecture can be implemented with a
similar approach, where one bit line contains devices of the group $G_+$ and another those of the group $G_-$. The total
synaptic current can then be found by subtracting the current of these two bit lines. To alter the synaptic weight, one of the word lines is activated according to the value of the selection counter to program the selected device. The scheme can also be adapted to alter the
weights of multiple synapses in parallel within the constraints of the maximum current that could flow through the bit line (see Supplementary Note
3).

\subsection{Multi-memristive synapses based on PCM devices} \label{subsecB}
\begin{figure*}[t!]
\centering
\begin{tabular}{c}
\includegraphics[scale = 1]{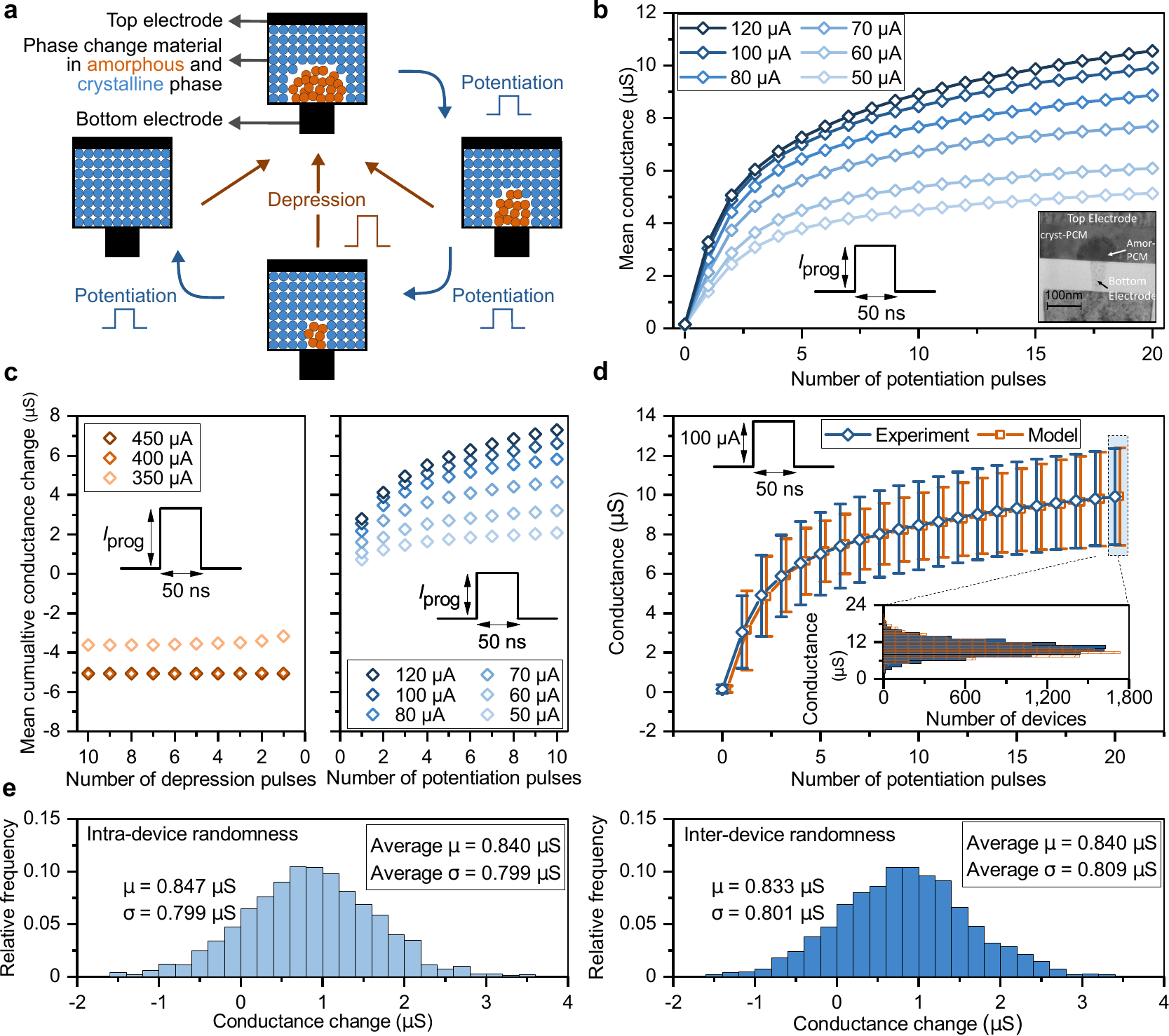}
\end{tabular}
\caption{\textbf{Synapses based on phase change memory.} (a) A PCM device consists of a phase-change material layer sandwiched
between top and bottom electrodes. The crystalline region can gradually be increased by the application of potentiation pulses. A depression pulse
creates an amorphous region that results in an abrupt drop in conductance, irrespective of the original state of the device. (b) Evolution of mean
conductance as a function of the number of pulses for different programming current amplitudes (\Iprog). Each curve is obtained by averaging the
conductance measurements from 9,700 devices. The inset shows a transmission electron micrograph of a characteristic PCM device used in this study.
(c) Mean cumulative conductance change observed upon the application of repeated potentiation and depression pulses. The initial conductance of the
devices is approximately \unit[5]{$\upmu$S}. (d) The mean and the standard deviation (1$\sigma$) of the conductance values as a function of number of pulses for
\Iprog $=$ \unit[100]{$\upmu$A} measured for 9,700 devices and the corresponding model response for the same number of devices. The distribution of conductance after the 20th potentiation
pulse and the corresponding distribution obtained with the model are shown in the inset. (e) The left panel shows a representative distribution of
the conductance change induced by a single pulse applied at the same PCM device 1,000 times. The pulse is applied as the 4th potentiation pulse to
the device. The same measurement was repeated on 1,000 different PCM devices, and the mean ($\upmu$) and standard deviation ($\sigma$) averaged over
the 1,000 devices are shown in the inset. The right panel shows a representative distribution of one conductance change induced by a single pulse on
1,000 devices. The pulse is applied as the 4th potentiation pulse to the devices. The same measurement was repeated for 1,000 conductance changes, and
the mean and standard deviation averaged over the 1,000 conductance changes are shown in the inset. It can be seen that the inter- and the intra-device
variability are comparable. The negative conductance changes are attributed to drift variability (see Supplementary Note 4).} \label{fig:exppcm}
\end{figure*}

In this section, we will demonstrate the concept of multi-memristive synapses using nanoscale PCM devices. A PCM device consists of a layer of phase change
material sandwiched between two metal electrodes (Fig. \ref{fig:exppcm}(a)) \cite{Y2016burrJETCAS}, which can be in a high-conductance crystalline phase or in a low-conductance amorphous phase. In an as-fabricated device, the material is typically in the crystalline
phase. When a current pulse of sufficiently high amplitude (referred to as the depression pulse) is applied, a significant portion of the
phase change material melts owing to Joule heating. If the pulse is interrupted abruptly, the molten material quenches into the amorphous phase as a
result of the glass transition. To increase the conductance of the device, a current pulse (referred to as the potentiation pulse) is
applied such that the temperature reached via Joule heating is above the crystallization temperature but below the melting point, resulting in the
recrystallization of part of the amorphous region \cite{Y2014sebastianNatComm}. The extent of crystallization depends on the amplitude and duration
of the potentiation pulse as well as on the number of such pulses. By progressively crystallizing the amorphous region by applying potentiation pulses, a continuum of conductance levels can be realized.

First, we present an experimental characterization of single-device PCM-based synapses based on doped \GST \ (GST) and integrated into a prototype chip
in \unit[90]{nm} CMOS technology \cite{Y2010closeIEDM} (see Methods). Figure \ref{fig:exppcm}(b) shows the evolution of the mean device conductance
as a function of the number of potentiation pulses applied. A total of 9,700 devices were used for the characterization, and the programming pulse
amplitude \Iprog \ was varied from \unit[50]{$\upmu$A} to \unit[120]{$\upmu$A}. It can be seen that the mean conductance value increases as a
function of the number of potentiation pulses. The dynamic range of conductance response is limited as the change in the mean conductance decreases
and eventually saturates with increasing number of potentiation pulses. Figure \ref{fig:exppcm}(c) shows the mean cumulative change in conductance as a
function of the number of pulses for different values of \Iprog. A well-defined nonlinear monotonic relationship exists between the mean cumulative
conductance change and the number of potentiation pulses. In addition, there is a granularity that is determined by how small a conductance change
can be induced by applying a single potentiation pulse. Large conductance change granularities as well as nonlinear conductance responses, both observed
in the PCM characterization performed here, have been shown to degrade the performance of neural networks trained with memristive synapses
\cite{Y2015burrITED,Y2017boybatPRIME}. Moreover, when a conductance decrease is desired,  a single high-amplitude depression pulse applied to a PCM
device has an all-or-nothing effect that fully depresses the device conductance to (almost) \unit[0]{$\upmu$S}. Such a strongly asymmetric
conductance response is undesirable in memristive-device-based implementations of neural networks\cite{Y2016gokmenFNINS}, and this is a significant
challenge for PCM-based synapses. Depression pulses with smaller amplitude could be applied to achieve higher conductance values. However, unlike the
potentiation pulses, it is not possible to achieve a progressive depression by applying successive depression pulses.

There are also significant intra- and inter-device variabilities associated with the conductance response in PCM devices as evidenced by the
distribution of conductance values upon application of successive potentiation pulses (see Fig. \ref{fig:exppcm}(d)). Note that the variability
observed in these devices fabricated in the 90 nm technology node is also found to be higher than that of those fabricated in the 180 nm node as reported
elsewhere \cite{Y2015burrITED}. Both the mean and variance associated with the conductance change depend on the mean conductance
value of the devices. We capture this behavior in a PCM conductance response model that relies on piece-wise linear approximations to the functions
that link the mean and variance of the conductance change to the mean conductance value \cite{Y2017nandakumarDRC}. As shown in Fig. \ref{fig:exppcm}(d),
this model approximates the experimental behavior fairly well.

The intra-device variability in PCM is attributed to the differences in atomic configurations associated with the amorphous phase change material created during the melt-quench process \cite{Y2016legalloESSDERC}. Inter-device variability, on the other hand, arises predominantly from the variability associated with the fabrication process across the array and results in significant differences in the maximum conductance and conductance response across devices (see Supplementary Fig. 1). To investigate the intra-device variability, we measured the conductance change on the same PCM device induced by a single potentiation pulse of amplitude \Iprog = \unit[100]{$\upmu$A} over 1,000 trials (Fig. \ref{fig:exppcm}(e), left panel). To quantify the inter-device variability, we monitored the conductance change induced by a single potentiation pulse across the 1,000 devices (Fig. \ref{fig:exppcm}(e), right panel). These experiments show that the standard deviation of the conductance change due to intra-device variability is almost as large as that due to the inter-device variability. The finding that the randomness in the conductance change is to a large extent intrinsic to the physical characteristic of the device implies that improvements in the array-level variability will not necessarily be effective in reducing the randomness.

\begin{figure*}[t!]
\centering
\begin{tabular}{c}
\includegraphics[scale = 1]{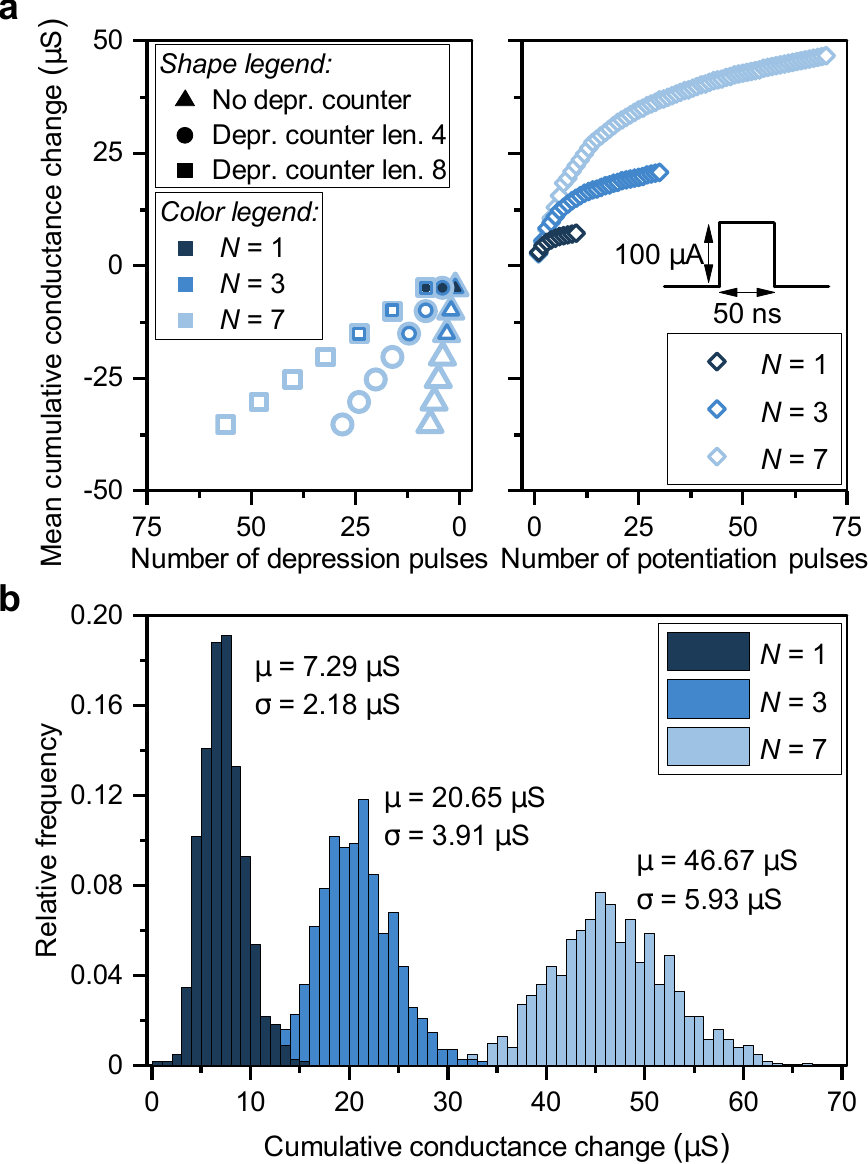}
\end{tabular}
\caption{\textbf{Multi-memristive synapses based on phase change memory.} (a) The mean cumulative conductance change is experimentally obtained for
synapses comprising 1, 3 and 7 PCM devices. The measurements are based on 1,000 synapses, whereby each individual device is initialized to a
conductance of approximately \unit[5]{$\upmu$S}. For potentiation, a programming pulse of \Iprog \ $=$ \unit[100]{$\upmu$A} was used, whereas for
depression,  a programming pulse of $I_\text{prog} =$ \unit[450]{$\upmu$A} was used. For depression, the conductance response can be made more
symmetric by adjusting the length of the depression counter. (b) Distribution of the cumulative conductance change after the application of 10, 30 and 70 potentiation pulses to 1, 3, and 7-PCM synapses, respectively. The mean ($\upmu$) and the variance ($\sigma ^2$) scale
almost linearly with the number of devices per synapse, leading to an improved weight update resolution.} \label{fig:expmulti}
\end{figure*}

The characterization work presented so far highlights the challenges associated with synaptic realizations using PCM devices and these can be generalized to other memristive technologies. The limited dynamic range, the asymmetric and nonlinear conductance response, the granularity and the randomness associated with conductance changes all pose challenges for realizing neural networks using memristive synapses. We now show how our concept of multi-memristive synapses can help in addressing some of those challenges. Experimental characterizations of multi-memristive synapses comprising 1, 3, and 7 PCM devices per synapse arranged in a non-differential architecture are shown in Fig. \ref{fig:expmulti}(a). The conductance change is averaged over 1,000 synapses. One selection counter with an increment rate of one arbitrates the device selection. As the total conductance is the sum of the individual conductance values, the dynamic range scales linearly with the number of devices per synapse. Alternatively, for a learning algorithm requiring a fixed dynamic range, multi-memristive synapses can improve the effective conductance change granularity. In addition, in contrast to a single device, the mean cumulative conductance change here is linear over an extended range of potentiation pulses. With multiple devices, we can also partially mitigate the challenge of an asymmetric conductance response. At any instance, only one device is depressed, which implies that the effective synaptic conductance decreases gradually in several steps instead of the abrupt decrease observed in a single device. Moreover, using the depression counter, the cumulative conductance changes for potentiation and depression can be made approximately symmetric by adjusting the frequency of depression events. Finally, Fig. \ref{fig:expmulti}(b) shows that both the mean and the variance of the conductance change scale linearly with the number of devices per synapse. Hence, the smallest achievable mean weight change decreases by a factor of $N$, whereas the standard deviation of the weight change decreases by $\sqrt{N}$, leading to an overall increase in weight update resolution by $\sqrt{N}$ (see Supplementary Fig. 2).

\subsection{Simulation results on handwritten digit classification}\label{sec:neuralNetSim}
In this section, we study the impact of PCM-based multi-memristive synapses in the context of training ANNs and SNNs. For synaptic potentiation, the PCM conductance response model presented above was used (see Fig. \ref{fig:exppcm}(d)). The depression pulses are assumed to cause an abrupt conductance drop to zero in a deterministic manner, modeling the PCM asymmetry. One selection counter is used for all synapses of the network, and the weight updates are done sequentially through all synapses in the same order at every pass. Potentiation and depression counters are used to balance the frequency of potentiation and depression events for $N > 1$.

\begin{figure*}[t!]
\centering
\begin{tabular}{c}
\includegraphics[scale = 1]{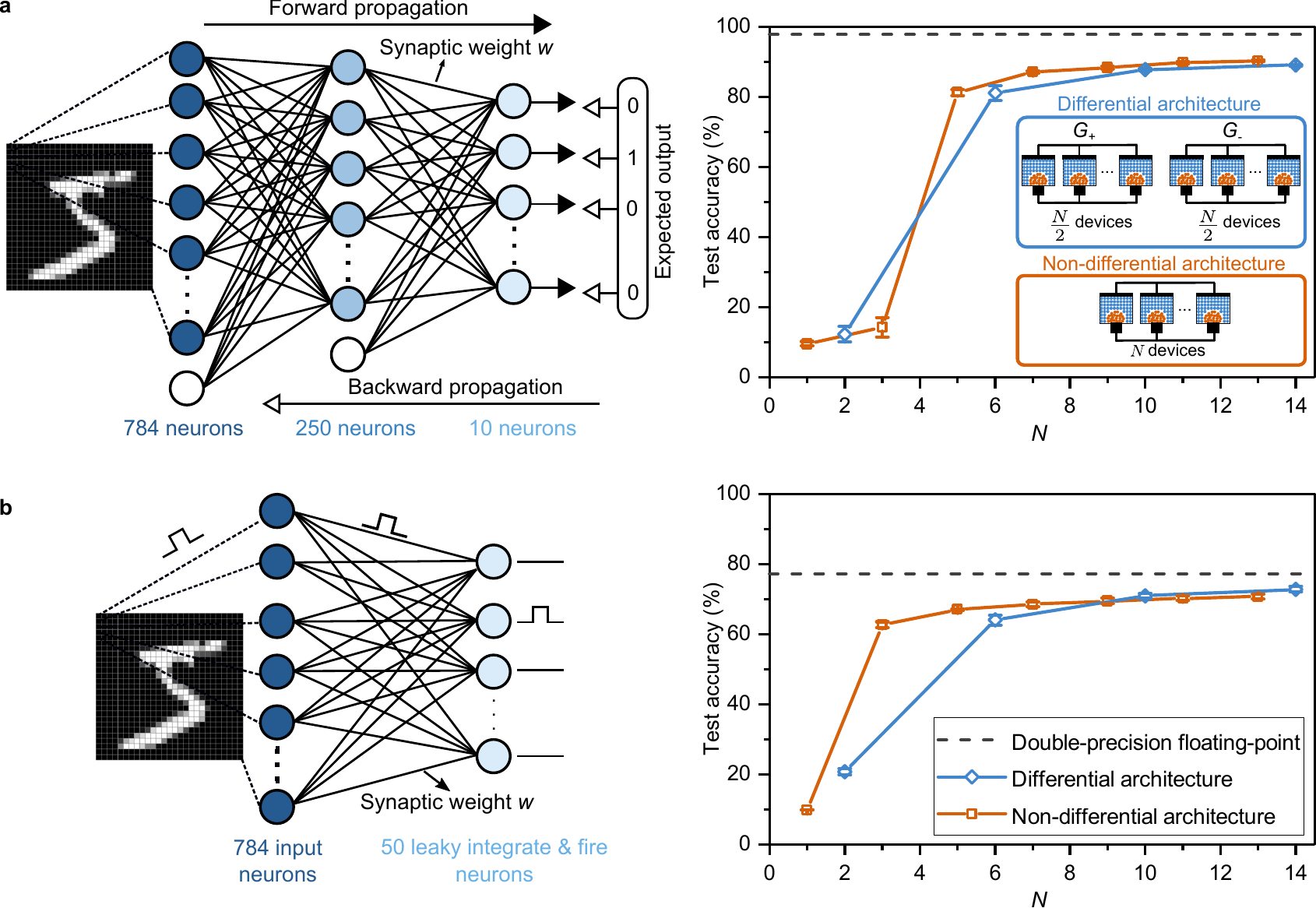}
\end{tabular}
\caption{\textbf{Applications of multi-memristive synapses in neural networks.} (a) An artificial neural network is trained using backpropagation to
perform handwritten digit classification. Bias neurons are used for the input and hidden neuron layers (white). A multi-memristive synapse model
based on the nonlinear conductance response of PCM devices is used to represent the synaptic weights in these simulations. Increasing the number of
devices in multi-memristive synapses (both in the differential and the non-differential architecture) improves the test accuracy. Simulations are
repeated for 5 different weight initializations. The error bars represent the standard deviation (1$\sigma$). The dotted line shows the test accuracy obtained
from a double-precision floating-point software implementation. (b) A spiking neural network is trained using an STDP-based learning rule for
handwritten digit classification. Here again, a multi-memristive synapse model is used to represent the synaptic weights in simulations where the
devices are arranged in the differential or the non-differential architecture. The classification accuracy of the network increases with the number of devices per synapse. Simulations are repeated for 5 different weight initializations. The error bars represent the standard deviation (1$\sigma$). The
dotted line shows the test accuracy obtained from a double-precision floating-point implementation.} \label{fig:nn}
\end{figure*}

First, we present simulation results that show the performance of an ANN trained with multi-memristive synapses based on the nonlinear conductance response model of the PCM devices. The feedforward fully-connected network with 3 neuron layers is trained with the backpropagation algorithm to perform a classification task on the MNIST data set of handwritten digits \cite{Y1998lecunIEEE} (see Fig. \ref{fig:nn}(a) and Methods). The ideal classification performance of this network, assuming double-precision floating-point accuracy for the weights, is 97.8\%. The synaptic weights are represented using the conductance values of a multi-memristive synapse model. In the non-differential architecture, a depression counter is used to improve the asymmetric conductance response and a potentiation counter to reduce the frequency of the potentiation events. As shown in Fig. \ref{fig:nn}(a), the classification accuracy improves with the number of devices per synapse. With the conventional differential architecture with 2 devices, the classification accuracy is below 15\%. With multi-memristive synapses in the differential architecture, we can achieve test accuracies exceeding 88.9\%, a performance better than the state-of-the-art in-situ learning experiments on PCM despite a significantly more nonlinear and stochastic conductance response due to technology scaling \cite{Y2015burrITED}. Remarkably, accuracies exceeding 90\% are possible even with the non-differential architecture, which clearly illustrates the efficacy of the proposed scheme.

In a second investigation, we studied an SNN with multi-memristive synapses to perform the same task of digit recognition, but with unsupervised learning \cite{Y2011querliozIJCNN} (see Fig. 4(b) and Methods). The weight updates are performed using an STDP rule: the synapse is potentiated whenever a presynaptic neuronal spike appears prior to a postsynaptic neuronal spike, and depressed otherwise. The amount of weight increase (decrease) within the potentiation (depression) window is constant and independent of the timing difference between the spikes. This necessitates a certain weight update granularity, which can be achieved by the proposed approach. The classification performance of the network trained with this rule using double-precision floating-point accuracy for the network parameters is 77.2\%. A potentiation counter is used to reduce the frequency of the potentiation events in both the differential and non-differential architectures, and a depression counter is used in the non-differential architecture to improve the asymmetric conductance response. The network could classify more than 70\% of the digits correctly for $N >$ 9 with both the differential and the non-differential architecture, whereas the network with the conventional differential architecture with 2 devices has a classification accuracy below 21\%.

In both cases, we see that the multi-memristive synapse significantly outperforms the conventional differential architecture with 2 devices, clearly illustrating the effectiveness of the proposed architecture. Moreover, the fact that the non-differential architecture achieves a comparable performance to that of the differential architecture is promising for synaptic realizations using highly asymmetric devices. A non-differential architecture would have a lower implementation complexity than its differential counterpart because the refresh operation \cite{Y2011suriIEDM, Y2015burrITED}, which requires reading and reprogramming $G_+$ and $G_-$, can be completely avoided.

\subsection{Experimental results on temporal correlation detection}\label{sec:SNN}
Next, we present an experimental demonstration of the multi-memristive synapse architecture using our prototype PCM chip (see Methods) to train an SNN that detects temporal correlations in event-based data streams in an unsupervised way. Unsupervised learning is widely perceived as a key computational task in neuromorphic processing of big data. It becomes increasingly important given today's variety of big data sources, for which often neither labeled samples nor reliable training sets are available. The key task of unsupervised learning is to reveal the statistical features of big data, and thereby shed light on its internal correlations. In this respect, detecting temporal and spatial correlations in the data is essential.

\begin{figure*}[t!]
\centering
\begin{tabular}{c}
\includegraphics[scale = 1]{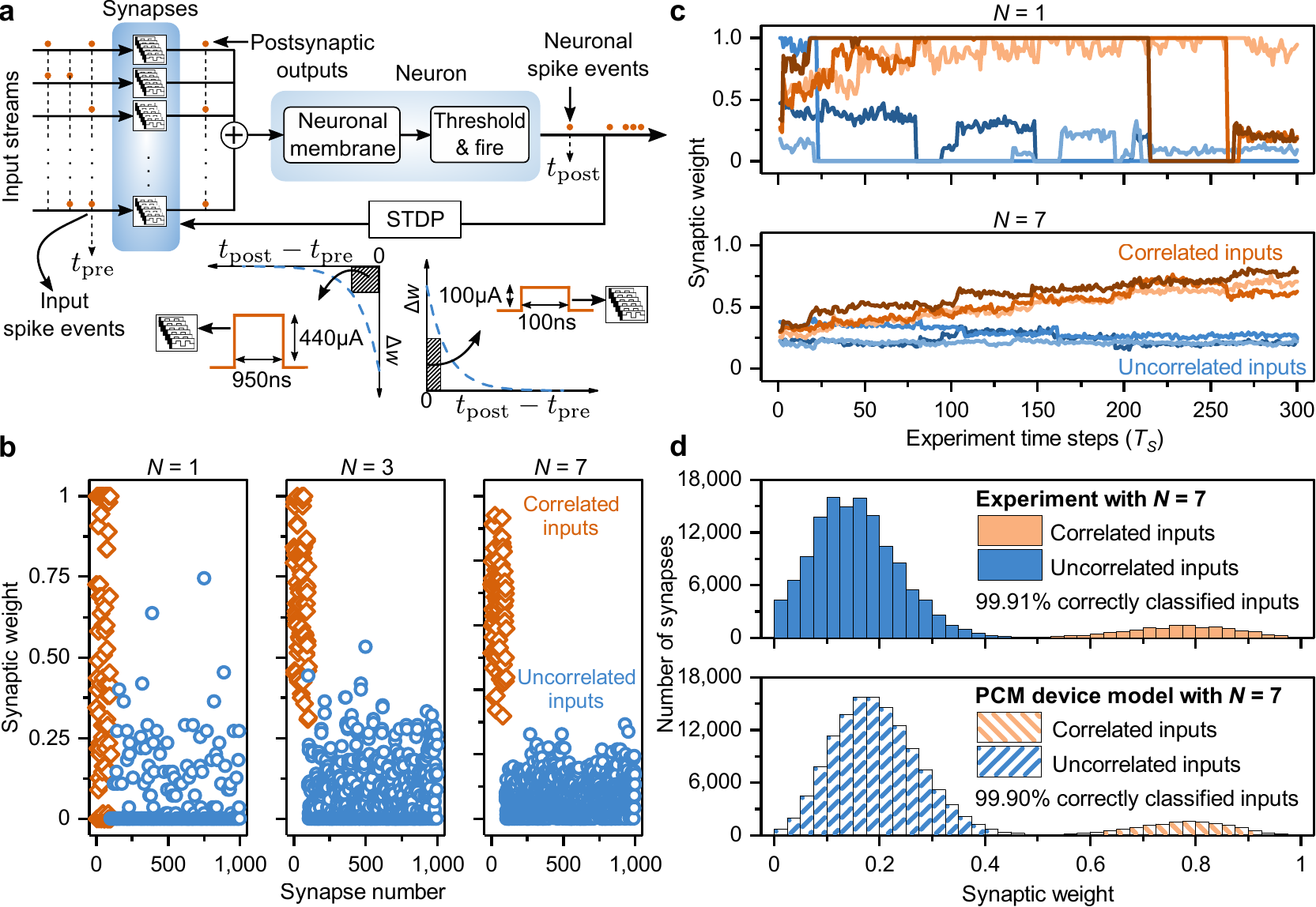}
\end{tabular}
\caption{\textbf{Experimental demonstration of multi-memristive synapses used in a spiking neural network.} (a) A spiking neural network is trained
to perform the task of temporal correlation detection through unsupervised learning. Our network consists of 1,000 multi-PCM synapses (in hardware)
connected to one integrate-and-fire (I\&F) software neuron. The synapses receive event-based data streams generated with Poisson distributions as
presynaptic input spikes. 100 of the synapses receive correlated data streams with a correlation coefficient of 0.75, whereas the rest of the
synapses receive uncorrelated data streams. The correlated and the uncorrelated data streams both have the same rate. The resulting postsynaptic
outputs are accumulated at the neuronal membrane. The neuron fires, i.e., sends an output spike, if the membrane potential exceeds a threshold. The
weight update amount is calculated using an exponential STDP rule based on the timing of the input spikes and the neuronal spikes. A potentiation
(depression) pulse with fixed amplitude is applied if the desired weight change is higher (lower) than a threshold. (b) The synaptic weights are
shown for synapses comprising $N = $1, 3 and 7 PCM devices at the end of the experiment (5,000 time steps). It can be seen that the weights of the
synapses receiving correlated inputs tend to be larger than the weights of those receiving uncorrelated inputs. The weight distribution shows
a clearer separation with increasing $N$. (c) Weight evolution of 6 synapses in the first 300 time steps of the experiment. The weight
evolves more gradually with the number of devices per synapse. (d) Synaptic weight distribution of an SNN comprising 144,000 multi-PCM
synapses with $N=7$ PCM devices at the end of an experiment (3,000 time steps) (upper panel). 14,400 synapses receive correlated input data
streams with a correlation coefficient of 0.75. A total of 1,008,000 PCM devices are used for this large-scale experiment. The lower panel shows the synaptic weight
distribution predicted by the PCM device model.} \label{fig:corr}
\end{figure*}

The SNN comprises a neuron interfaced to plastic synapses, with each one receiving an event-based data stream as presynaptic input spikes \cite{Y2003gutigJNeuro, Y2016tumaNatNano} (see Fig. \ref{fig:corr}(a) and Methods).  A subset of the data streams are mutually temporally correlated, whereas the rest are uncorrelated (see Supplementary Note 5). When the input streams are applied, postsynaptic outputs are generated at the synapses that received a spike. The resulting postsynaptic outputs are accumulated at the neuron. When the neuronal membrane potential exceeds a threshold, the output neuron fires, generating a spike. The synaptic weights are updated using an STDP rule; synapses that receive an input spike within a time window before (after) the neuronal spike get potentiated (depressed). As it is more likely that the temporally correlated inputs will eventually govern the neuronal firing events, the conductance of synapses receiving correlated inputs is expected to increase, whereas that of synapses whose input are uncorrelated is expected to decrease. Hence, the final steady-state distribution of the weights should display a separation between synapses receiving correlated and uncorrelated inputs.

First, we perform small-scale experiments in which multi-memristive synapses with PCM devices are used to store the synaptic weights. The network comprises 1,000 synapses, of which only 100 receive temporally correlated inputs with a correlation coefficient $c$ of 0.75. The difficulty in detecting whether an input is correlated or not increases both with decreasing $c$ and decreasing number of correlated inputs. Hence, detecting only 10\% correlated inputs with $c<$1 is a fairly difficult task and requires precise synaptic weight changes for the network to be trained effectively \cite{Y2017sebastianNatComm}. Each synapse comprises $N$ PCM devices organized in a non-differential architecture. During the weight update of a synapse, a single potentiation pulse or a single depression pulse is applied to one of the devices the selection counter points to. A depression counter with a maximum value of two is incorporated for $N > $ 1 to balance the PCM asymmetry. Figure \ref{fig:corr}(b) depicts the synaptic weights at the end of the experiment for different values of $N$. To quantify the separation of the weights receiving correlated and uncorrelated inputs, we set a threshold weight that leads to the lowest number of misclassifications. The number of misclassified inputs were 49, 8 and 0 for $N = $1, 3 and 7 respectively. This demonstrates that the network's ability to detect temporal correlations increases with the number of devices. This holds true even for lower values of the correlation coefficient as shown in Supplementary Note 6. With $N=1$, there are strong abrupt fluctuations in the evolution of the conductance values because of the abrupt depression events as shown in Fig. \ref{fig:corr}(c). With $N = 7$, a more gradual potentiation and depression behavior is observed. For $N=7$, the synapses receiving correlated and uncorrelated inputs can be perfectly separated at the end of the experiments. In contrast, the weights of correlated inputs display a wider weight distribution and there are numerous misclassified weights for $N=1$.

The multi-memristive synapse architecture is also scalable to larger network sizes. To demonstrate this, we repeated the above correlation experiment with 144,000 input streams, and with 7 PCM devices per synapse, resulting in more than one million PCM devices in the network. As shown in Fig. \ref{fig:corr}(d), well-separated synaptic distributions have been achieved in the network at the end of the experiment. Moreover, a simulation was performed with the nonlinear PCM device model (see Methods). The simulation captures the separation of weights receiving correlated and uncorrelated inputs. In both experiment and simulation, approximately 0.1\% of the inputs were misclassified after training.

\section{Discussion}
The proposed synaptic architecture bears similarities to several aspects of neural connectivity in biology, as biological neural connections also comprise multiple sub-units. For instance, in the central nervous system, a presynaptic neuron may form multiple synaptic terminals (so-called boutons) to connect to a single postsynaptic neuron \cite{Y1991NatureEdwards}. Moreover, each biological synapse contains a plurality of presynaptic release sites \cite{Y1997bolshakovNeuron} and postsynaptic ion channels \cite{Y1999malenkaScience}. Furthermore, our implementation of plasticity through changes in the individual memristors is analogous to individual plasticity of the synaptic connections between a pair of biological neurons \cite{Y2004malenkaNeuron}, which is also true for the individual ion channels of a synaptic connection \cite{Y1998benkeNature, Y2004malenkaNeuron}. The involvement of progressively larger numbers of memristive devices during potentiation is analogous to the development of new ion channels in a potentiated synapse \cite{Y1997bolshakovNeuron, Y1999malenkaScience}.

A significant advantage of the proposed multi-memristive synapse is its crossbar compatibility. In memristive crossbar arrays, matrix-vector multiplications associated with the synaptic efficacy can be implemented with a read operation achieving \textit{O}(1) complexity. Note that memristive devices can be read with low energy (\unit[10]-\unit[100]{fJ} for our devices), which leads to a substantially lower energy consumption than in conventional von Neumann systems \cite{Y2017legalloIEDM, Y2018liNatElectronics, Y2018legalloNatElectronics}. Furthermore, the synaptic plasticity is realized in place without having to read back the synaptic weights. Even though, the power dissipation associated with programming the memristive devices is at least ten times higher than that required for the read operation, as only one device of the multi-memristive synapse is programmed at each instance of synaptic update, our scheme does not introduce a significant energy overhead. Memristive crossbars can also be fabricated with very small areal footprint \cite{Y2013alibartNatComm, Y2014eryilmazFNINS, Y2015preziosoNature}. The neuron circuitry of the crossbar array, which typically consumes a larger area than the crossbar array, only increases marginally owing to the additional circuitry needed for arbitration. Finally, because even a single global counter can be used for arbitrating a whole array, the additional area / power overhead is expected to be minimal.

The proposed architecture also offers several advantages in terms of reliability. The other constituent devices of a synapse could compensate for the occasional device failure. In addition, each device in a synapse gets programmed less frequently than if a single device were used, which effectively increases the overall lifetime of a multi-memristive synapse. The potentiation and depression counters reduce the effective number of programming operations of a synapse, further improving endurance-related issues.

Device selection in the multi-memristive synapse is performed based on the arbitration module alone, without knowledge of the conductance values of the individual devices, thus there is a non-zero probability that a potentiation (depression) pulse will not result in an actual potentiation (depression) of the synapse. This would effectively translate into a weight-dependent plasticity whereby the probability to potentiate reduces with increasing synaptic weight and the probability to depress reduces with decreasing synaptic weight (see Supplementary Notes 7, 8). This attribute could affect the overall performance of a neural network. For example, weight-dependent plasticity has been shown to impact the classification accuracy negatively in an ANN \cite{Y2016sidlerESSDERC}. In contrast, a study suggests that it can stabilize an SNN intended to detect temporal correlations \cite{Y2003gutigJNeuro}.

The ANN and SNN simulations in Section \ref{sec:neuralNetSim} with the PCM model perform worse, even in the presence of multi-memristive synapses with $N >$ 10, than the simulations with double-precision floating-point weights. We show that this behavior does not arise from the weight-dependent plasticity of the multi-memristive synapse scheme, but from the nonlinear PCM conductance response (see Supplementary Fig. 9). Using a uni-directional linear device model where the conductance change is modeled as a Gaussian random number with mean (granularity) and standard deviation (stochasticity) of \unit[0.5]{$\upmu$S}, accuracies exceeding 96.7\% are possible in ANN with only 1\% performance loss compared with double-precision floating-point weights. Similarly, the network can classify more than 77\% of the digits correctly in the SNN using the linear device model, reaching the accuracy of the double-precision floating-point weights.

Note also that the drift in conductance states, which is unique to PCM technology, does not appear to have a significant impact on our studies. As described recently \cite{Y2016fumarolaICRC}, as long as the drift exponent is small enough ($<0.1$; in our devices it is on average 0.05, see Supplementary Note 4), it is not very detrimental for neural network applications. Our own experimental results on SNNs presented in Section \ref{sec:SNN} point in this direction, as the network seems to maintain the classification accuracy despite drift. Although conductance drift is not intended to be countered using the multi-memristive concept, there are attempts to overcome it via advanced device-level ideas \cite{Y2015koelmansNatComm}, which could be used in conjunction with a multi-memristive synapse.

In the presence of significant nonlinear conductance response and drift, one could envisage an alternate multi-memristive synaptic architecture in which multiple devices are used to store the weights, but with varying significance. For instance, if $N$-bit synaptic resolution is required, $N$ memory devices could be used, with each device programmed to the maximum (fully potentiated) or minimum (fully depressed) conductance states to represent a number in binary format. In such a binary system, for synaptic efficacy, each device needs to be read independently, which could be accomplished by reading each of the $N$ bits one by one, or alternatively, $N$  amplifiers could be used to read the $N$ bits in parallel. For synaptic plasticity, the desired weight update should be done with prior knowledge of the stored weight. Otherwise, a blind update could have a large detrimental effect, especially if the error is associated with devices representing the most significant bits. However, a direct comparison between these alternate architectures and our proposed scheme requires a detailed system-level investigation which is beyond the scope of this paper.

In summary, we propose a novel synaptic architecture comprising multiple memristive devices with non-ideal characteristics to efficiently implement learning in neural networks. This architecture is shown to overcome several significant challenges that are characteristic to nanoscale memristive devices proposed for synaptic implementation, such as the asymmetric conductance response, limitations in resolution and dynamic range, as well as device-level variability. The architecture is applicable to a wide range of neural networks and memristive technologies and is crossbar-compatible. The high potential of the concept is demonstrated experimentally in a large-scale SNN performing unsupervised learning. The proposed architecture and its experimental demonstration are a significant step towards the realization of highly efficient, large-scale neural networks based on memristive devices with typical, experimentally observed non-ideal characteristics.

\newpage
\section{Methods}
\subsection{Experimental platform}
The experimental hardware platform is built around a prototype PCM chip with 3 million devices with a 4-bank inter-leaved architecture. The
mushroom-type PCM devices are based on doped \GST \ (GST) and were integrated into the prototype chip in \unit[90]{nm} CMOS technology, based on an
existing fabrication recipe \cite{Y2010closeIEDM}. The radius of the bottom electrode is approximately \unit[20]{nm}, and the thickness of the phase
change material is approximately \unit[100]{nm}.  A thin oxide n-type field-effect transistor (FET) enables access to each PCM device. The chip also
integrates the circuitry for addressing, an 8-bit on-chip analog-to-digital converter (ADC) for readout, and voltage- or current-mode programming. An
analog-front-end (AFE) board is connected to the chip and accommodates digital-to-analog converters (DACs) and ADCs, discrete electronics, such as
power supplies, voltage and current reference sources. An FPGA board with embedded processor and Ethernet connection implements the overall system
control and data management.

\subsection{PCM characterization}
For the experiment of Fig. \ref{fig:exppcm}(b), measurements were done on 10,000 devices. All devices were initialized to approximately \unit[0.1]{$\upmu$S} with an iterative procedure. In the experiment, 20 potentiation pulses with a duration of \unit[50]{ns} and varying amplitudes were applied. After each potentiation pulse, the devices were read 50 times in approximately 5 s intervals. The reported device conductance for a potentiation pulse is the average conductance obtained by the 50 consecutive read operations. This method is used to minimize the impact of drift \cite{Y2015sebastianIRPS} and read noise \cite{Y2010closeIEDM}. At the end of the experiment, approximately 300 devices were excluded because they had an initial conductance of less than \unit[0.1]{$\upmu$S} or a final conductance after 20 potentiation pulses of more than \unit[30]{$\upmu$S}.

In the measurements for Fig. \ref{fig:exppcm}(c), 10,000 devices were used. The data was obtained after initializing the device conductances to \unit[5]{$\upmu$S} by an iterative procedure. Next, potentiation (depression) pulses of varying amplitude and \unit[50]{ns} duration were applied. Every potentiation (depression) pulse was followed by 50 read operations done approximately 5 s apart. The device conductance was averaged for the 50 read operations.

In the experiments of Fig. \ref{fig:exppcm}(e), 1,000 devices were used. All devices were initialized to approximately \unit[0.1]{$\upmu$S} with an iterative procedure. This was followed by 4 potentiation pulses of amplitude \Iprog \ = \unit[100]{$\upmu$A} and width \unit[50]{ns}. After the last 2 potentiation pulses, devices were read 20 times with the reads approximately 1.5 s apart. The device conductances for 20 read operations were averaged. The difference between the averaged conductances for the 3rd and 4th potentiation pulses is defined as the conductance change. This experimental sequence was repeated on the same devices for 1,000 times so that 1,000 conductance changes were measured for each device.

For the experiments of Fig. \ref{fig:expmulti}, measurements were done on 1,000, 3,000, and 7,000 devices for $N = 1, 3 \text{ and } 7$, respectively. Device conductances were initialized to \unit[5]{$\upmu$S} by an iterative procedure. Next, for potentiation, programming pulses of amplitude \unit[100]{$\upmu$A} and width \unit[50]{ns} were applied. For depression, programming pulses of \unit[450]{$\upmu$A} amplitude and \unit[50]{ns} width were applied. After each potentation (depression) pulse, device conductances were read 50 times and averaged. The delay between each read event was approximately \unit[5]{s}.

In all measurements, device conductances were obtained by applying a fixed voltage of \unit[0.3]{V} amplitude and measuring the corresponding current.

\subsection{Simulation of neural networks}
The ANN contains 784 input neurons, 250 hidden layer neurons, and 10 output neurons. In addition, there is one bias neuron at the input layer and one
bias neuron at the hidden layer. For training, all 60,000 images from the MNIST training set are used in the order they appear in the database over
10 epochs. Subsequently, all 10,000 images from the MNIST test set are shown for testing. The test set is applied at every 1,000th example for the last 20,000
images of the 10th epoch of training, and the results are averaged. The input images from the MNIST set are greyscale pixels with values ranging from 0 to 255 and have a size of 28
times 28. Each of the input layer neurons receives input from one image pixel, and the input is the pixel intensity scaled by 255 in double-precision floating-point. The neurons of the hidden and the output layers are sigmoid neurons. Synapses are multi-memristive, and each synapse comprises
$N$ devices. The devices in a synapse are arranged using either a non-differential or a differential architecture. In the non-differential
architecture, we scale the device conductance of  \unit[0]{$\upmu$S} to weight $-\frac{1}{N}$ and that of \unit[10]{$\upmu$S} to weight $\frac{1}{N}$. The weight is not incremented further if it exceeds $\frac{1}{N}$ to model the PCM saturation behavior. The minimum weight is $-\frac{1}{N}$ because the minimum device conductance is \unit[0]{$\upmu$S}. The weight of each device $w_n$ is initialized randomly with a uniform distribution in the interval $[\frac{-1}{2N}, \frac{1}{2N}]$. The total synaptic weight
is calculated as $\sum_{n=1}^{N} w_n$. In the differential architecture, $N$ devices are arranged in two sets, where $\frac{N}{2}$ devices represent
$G_+$ and $\frac{N}{2}$ devices represent $G_-$. We scale the device conductance of \unit[0]{$\upmu$S} to weight 0 and that of \unit[10]{$\upmu$S} to weight
$\frac{2}{N}$. The weight is not incremented if it exceeds $\frac{2}{N}$ and the minimum weight is 0. The weight of each device $w_{n+,n-}$ for $n =$ 1, 2, ..., $\frac{N}{2}$ is
initialized randomly with a uniform distribution in the interval $[\frac{1}{N}, \frac{2}{N}]$.  The total synaptic weight $w$ is $(\sum_{n} w_{n+}) -
(\sum_{n} w_{n-})$. For double-precision floating-point simulations, the synaptic weights are initialized with a uniform distribution in the interval [-0.5,
0.5]. The weight updates $\Delta w$ are done sequentially to synapses, and the selection counter is incremented by one after each weight update. If
$\Delta w >0$, the synapse will undergo potentiation. In both architectures, each potentiation pulse on average would induce a weight change of size
$\epsilon = \frac{0.1}{N}$ if a linear model was used; the number of potentiation pulses to be applied are calculated by rounding $\frac{\Delta
w}{\epsilon}$. Then, for each potentiation pulse, an independent Gaussian random number with mean and standard deviation according to the model of Fig.
\ref{fig:exppcm}(d) is added. This weight change is applied to the device to which the selection counter points. If $\Delta w <0$, the synapse will
undergo a depression. In the differential architecture, a potentiation pulse is applied to a device from the set representing $G_-$ using the
above-mentioned methodology. In the non-differential architecture, a depression pulse is applied to one of the devices pointed at by the selection
counter if $\Delta w < 0.5\epsilon$. The weight of the device drops to 0. For $N > 1$, we used a depression counter of length 5 and a potentiation
counter of length 2. No depression or potentiation counter is used for $N = 1$. In the differential architecture, after the weight change has been
applied for potentiation and depression, synapses are checked for the refresh operation. If there is a synapse which has $w_+ > 0.9$ or $w_- > 0.9$,
a refresh is done on that synapse; $w$ is recorded, and all devices in the synapse are set to 0. The programming will be done to devices of the set
$w_+$ if $w>0$  or to devices of the set $w_-$ if $w<0$. The number of potentiation pulses is calculated by rounding $\frac{\Delta w}{\epsilon}$.
The pulses are applied to all devices, starting with the first device of the set. One independent Gaussian random number with mean and standard
deviation according to the model of Fig. \ref{fig:exppcm}(d) is calculated for each of the potentiation pulses. The learning rate is 0.4 for all
simulations.

The SNN comprises 784 input neurons and 50 output neurons. These synapses are multi-memristive, and each synapses consists of $N$ memristive
devices. The network is trained with all 60,000 images from the MNIST set over 3 epochs and tested with all 10,000 test images from the set. The test
set is applied at every 1,000th example for the last 20,000 images, and the results are averaged. The simulation time step is \unit[5]{ms}. Each input
neuron receives input from one pixel of the input image. Each input image is presented for \unit[350]{ms}, and the information regarding the
intensity of each pixel is in the form of spikes. We create the input spikes using a Poisson distribution, where independent Bernoulli trials are
conducted to determine whether there is a spike at a time step.  A spike rate is calculated as $\frac{\text{pixel intensity}}{255} \times  20 \
\text{Hz}$. A spike is generated if $(\text{spike rate}  \times  5 \ \text{ms} > x)$, where $x$ is a uniformly distributed random number between 0
and 1. The input spikes create a current with the shape of a delta function at the corresponding synapse. The magnitude of this current is equal to
the weight of the synapse. The synaptic weights $w$ are learned with an SDTP rule \cite{Y2011querliozIJCNN}. The synapses are arranged in a
non-differential or a differential architecture. In both architectures, we scale the device conductance of \unit[0]{$\upmu$S} to weight 0 and
that of \unit[10]{$\upmu$S} to weight $\frac{1}{N}$. The weight is not incremented further if it exceeds $\frac{1}{N}$. The minimum weight is 0 because the
minimum device conductance is \unit[0]{$\upmu$S}. In the non-differential architecture, the weight of each device $w_n$ is initialized randomly with
a uniform distribution in the interval $[\frac{2}{5N}, \frac{3}{5N}]$. The total synaptic weight is calculated as $\sum_{n=1}^{N} w_n$. In the
differential architecture, $N$ devices are arranged in two sets. The weight of each device $w_{n+,n-}$ for $n =$ 1, 2, ..., $\frac{N}{2}$ is
initialized randomly with a uniform distribution in the interval $[\frac{3}{5N}, \frac{4}{5N}]$.  The total synaptic weight is $(\sum_{n} w_{n+}) -
(\sum_{n} w_{n-}) + 0.5$. For double-precision floating-point simulations, the synaptic weights are initialized with a uniform distribution in the interval
[0.25, 0.75].  At each simulation time step, the synaptic currents are summed at the output neurons and accumulated using a state variable $X$. The
output neurons are of the leaky integrate-and-fire type and have a leak constant of $\tau = $ \unit[200]{ms}. Each output neuron has a spiking
threshold. This spiking threshold is set initially to 0.125 (note that the sum of the currents is normalized by the number of input neurons) and is
altered by homeostasis during training. An output neuron spikes when $X$ exceeds the neuron threshold. Only one output neuron is allowed to spike at
a single time step, and if the state variables of several neurons exceed their threshold, then the neuron whose state variable exceeds its threshold
the most is the winner. The state variables of all other neurons are set to 0 if there is a spiking output neuron. If there is a postsynaptic
neuronal spike, the synapses that received a presynaptic spike in the past \unit[30]{ms} are potentiated. If there is a presynaptic spike the
synapses that had a postsynaptic neuronal spike in the past \unit[1.05]{s} are depressed. The weight change amount is constant for potentiation
($\Delta w_+ =$ 0.01) and depression ($\Delta w_- =$ 0.006), following a rectangular STDP rule.  The weight updates are done using the scheme
described above with $\epsilon = \frac{0.05}{N}$. For the non-differential architecture, a depression pulse is applied when $\Delta w < 0$. The
depression counter length is set to the floor of $\frac{1}{N  \times  0.006}$ for $N > 1$. In the non-differential and the differential
architecture, a potentiation counter of length 3 and 2 is used, respectively. After the 1,000th input image, upon presentation of every two images,
the spiking thresholds of the output neurons are adjusted through homeostasis. The threshold increase for every output neuron is calculated as
$0.0005  \times  (\text{A} - \text{T})$, where $A$ is the activity of the neuron and $T$ is the target firing rate. $A$ is calculated as
$\frac{S}{350 \text{ms}  \times  100}$, where $S$ is the sum of the neuron's firing event in the past 100 examples. We define the $T$ as
$\frac{5}{350 \text{ms}  \times  50}$, where 50 is the number of output neurons in the network. After training, the synaptic weights and the neuron
thresholds are kept constant. To quantify how well the training is, we show all 60,000 images to the network, and the neuron that spikes the most
often during the presentation of an image for \unit[350]{ms} is recorded. The neuron is mapped to the class, i.e., to one of the 10 digits, for which
it spiked the most. This mapping is then used to detect the classification accuracy when the test set is presented.

\subsection{Correlation detection experiment}
The network for correlation detection comprises 1,000 plastic synapses connected to an output neuron. Each synapse is multi-memristive and consists
of $N$ devices. The synaptic weights $w \in [0, 1]$ are learned with an STDP rule \cite{Y2000songNatNeuro}. Because of the hardware latency, we will
use normalized units to describe time in the experiment. The experiment time steps are of size $T_\text{s} = $ 0.1. Each synapse receives a stream of
spikes, and the spikes have the shape of the delta function. 100 of the input spike stream are correlated. The correlated and the uncorrelated spike
streams have equal rates of $r_{\text{cor}} = r_{\text{uncor}} = $1. The correlated inputs share the outcome of a Bernoulli trial. This Bernoulli
trial is described as $B = x > 1 - r_{\text{cor}}  \times  T_\text{s}$, where $x$ is a uniformly distributed random number. By using this event, the input spikes
for the correlated streams are generated as $B  \times  (r_{\text{cor}}  \times  T_\text{s} + \sqrt{c}  \times  (1 - r_{\text{cor}}  \times  T_\text{s}) > x_1) + \sim B  \times  (r_{\text{cor}}  \times  T_\text{s}
 \times  (1 - \sqrt{c}) > x_2)$, where $x_1$ and $x_2$ are uniformly distributed random  numbers, $c$ is the correlation coefficient of value 0.75, and
$\sim$ denotes the negation operation \cite{Y2003gutigJNeuro, Y2017sebastianNatComm}.  The uncorrelated processes are generated as $x_3 > 1 -
(r_{\text{unco}r}  \times  T_\text{s})$, where $x_3$ is a uniformly distributed random variable.  Note that the probability of generating a spike is low because
$r_{\text{cor, uncor}}  \times  T_\text{s} \ll 1$. These input spikes generate a current of the size of the synaptic weights. At every time
step, the currents are summed and accumulated at the neuronal membrane variable $X$.  The neuronal firing events in any given time step are
determined only by the spiking events that occur in that time step. If $X$ exceeds a threshold of 52, the output neuron fires. The weight update
calculation follows an exponential STDP rule where the amount of potentiation is calculated as $A_+e^{-|\Delta t| / \tau_+}$ and that of depression is
calculated as $-A_-e^{-|\Delta t| / \tau_-}$. $A_+, A_-$ are the learning rates,  $\tau_+, \tau_-$ are time constants, and $\Delta t$ is the time
difference between the input spikes and the neuronal spikes. We set  $2  \times  A_+ = A_- = 0.004$ and  $\tau_+ = \tau_- = 3  \times  T_\text{s}$. The
higher-order pairs of spikes are also considered in our algorithm.

The weight storage and weight update operations are done on PCM devices. We access each PCM device sequentially for reading and programming. For
device initialization, an iterative procedure is used to program the device conductances to \unit[0.1]{$\upmu$S} and this is followed by one
potentiation pulse of amplitude \Iprog$ = $\unit[120]{$\upmu$A} and \unit[50]{ns} width. Although the weight update is calculated using an
exponential STDP rule, it is applied following a rectangular STDP rule. For potentiation, a single potentiation pulse of amplitude \Iprog \ $ =
$ \unit[100]{$\upmu$A} and \unit[100]{ns} width is applied when $\Delta w_+ \geq 0.001$. For depression, a single depression pulse of amplitude \Iprog \ 
$= $ \unit[440]{$\upmu$A} and \unit[950]{ns} width is applied when $\Delta w_- \leq -0.001$. The potentiation and depression pulses are sent to one
device from the multi-memristive synapse the selection counter points to. When applying depression pulses, a depression counter of length 2 is used
for $N > 1$. After each programming operation, the device conductances are read by applying a fixed voltage of amplitude \unit[0.2]{V} and measuring the corresponding current. The conductance value $G$ of a device is converted to its synaptic
weight as $w_n = \frac{G}{N  \times  \ 9.5 \upmu\text{S}}$. The weights of the devices in a multi-memristive synapse are summed to calculate the total
synaptic weight $\sum_{n=1}^{N} w_n $.

For the large-scale experiment, 144,000 synapses are trained, of which 14,400 receive correlated inputs. Each multi-memristive synapse comprises $N = $ 7 devices, and a total of 1,008,000 PCM devices are used for this experiment. The same network parameters as in the small-scale experiment are used,
except for the neuron threshold. The neuron threshold is scaled with the number of synapses and is set to 7,488. The learning algorithm and
conductance-to-weight conversion are identical to those in the small-scale experiment.

The nonlinear PCM model used for the accompanying simulation study is based on the conductance evolution of PCM devices with \Iprog$ = $\unit[100]{$\upmu$A} pulse amplitude and a pulse width of \unit[50]{ns}. Two potentiation pulses are applied consecutively to capture the
conductance change behavior of one potentiation pulse with pulse width \unit[100]{ns} of the experiments. A depression pulse is assumed to set the
device conductance to \unit[0]{$\upmu$S}, irrespective of the conductance value prior to the application of the pulse.

\subsection{Data availability}
The data that support the findings of this study are available from the corresponding authors upon request.

\newpage
\section*{References}
\nocite{*}
\bibliographystyle{naturemag}

\section*{Acknowledgments}
We would like to thank N. Papandreou, U. Egger, S. Wozniak, S. Sidler, A. Pantazi and G. Burr for technical input. I. B. and T. M. would like to
acknowledge financial support from the Swiss National Science Foundation. A. S. would like to acknowledge funding from the European Research Council
(ERC) under the European Union's Horizon 2020 research and innovation programme (grant agreement No. 682675).

\section*{Author contributions}
I.B., M. L. G., T. T. and A. S. designed the concept. I. B., N. S. R. and T. M. performed the simulations. I. B., N. S. R., M. L. G. and A. S.
performed the experiments. T. M and T. P. provided critical insights. I. B., M. L. G., and A. S. co-wrote the manuscript with input from the other
authors. A. S., Y. L., B. R. and E. E. supervised the work.

\section*{Competing interests}
The authors declare no competing interests.

\newpage

\end{document}